\documentclass[twocolumn,showpacs,showkeys, preprintnumbers,amsmath,amssymb]{revtex4}
\usepackage{graphicx}
\usepackage{dcolumn}
\usepackage{bm}
\usepackage{amsmath}
\newcommand{\beqn}{\begin{eqnarray}}
\newcommand{\eeqn}{\end{eqnarray}}

\begin{document}
\title{Two-dimensional relativistic hydrogenic atoms: A complete set of constants of motion}
\author{A. Poszwa}
\author{A. Rutkowski}

\affiliation{%
Department of Physics and Computer Methods, University of Warmia and Mazury in Olsztyn, ul. \.Zo{\l}nierska 14,
 10-561 Olsztyn, Poland.\\}%

\date{\today}
\begin{abstract}
{The complete set of operators commuting with the Dirac
Hamiltonian and exact analytic solution of the Dirac equation for
the two-dimensional Coulomb potential is presented. Beyond the
eigenvalue $\mu$ of the operator $j_{z}$, two quantum numbers
$\eta$ and $\kappa$ are introduced as eigenvalues of hermitian
operators $P=\beta\sigma'_{z}$ and
$K=\beta(\sigma'_{z}l_{z}+1/2)$, respectively. The classification
of states according to the full set of constants of motion without
referring to the non-relativistic limit is proposed. The linear
Paschen-Back effect is analyzed using exact field-free
wave-functions as a zero-order approximation.}
\end{abstract}


 \maketitle

\section{Introduction}
Low-dimensional quantum systems have been the focus of extensive
theoretical investigations in the last decades. Technological
advances in semiconductor physics and recent developments in
nanostructure technology provide techniques of creating
low-dimensional structures like superlatices, quantum dots,
quantum wires or quantum wells \cite{NS}. The most representative
analogues of hydrogenlike systems in the world of semiconductors
are hydrogenic donors being the bound states of conduction
electron and a donor impurity \cite{JZJ} and Wannier-Mott excitons
formed by an electron and a hole \cite{RC}. After the
renormalization of Coulomb potential by introducing of dielectric
constant and replacing the electron mass by effective one, the
\textit{atomic} objects in two-dimensional structures can be
treated as 2-D hydrogelike atoms. At this stage, the 2-D hydrogen
problem determines a leading approximation for study of hydrogen
type bound states in extreme anisotropic crystals, in which the
$z$-component of a diagonal anisotropic mass tensor is much larger
then the two remaining ones \cite{WZS}. The non- and
weak-relativistic \cite{VIL} approaches are usually considered as
sufficiently good approximations to the realistic description of the 2-D
objects in solid matter. However, in a searching of the
quantum-mechanical properties of the 2-D systems interacting with
low-dimensional gauge fields \cite{KAM, SHK}, the complete
relativistic theory is inevitable.

The quantum mechanical two-dimensional central problem, with the
Coulomb potential $-Z/\rho$ has been solved by many authors. In
the nonrelativistic theory the analytic solution can be derived in
strict analogy to the three-dimensional Coulomb problem, after the
separation of the Schr\"odinger equation in polar coordinates
\cite{YG1,ZAS} or in parabolic coordinates \cite{CIS}. The
solution of the relativistic 2-D hydrogen-like problem has been
obtained in the framework of the two-component approach \cite{DM}.
Although the formalism based on the two-dimensional representation
of gamma matrices gives correct formula for energy levels, it does
not provide a good background for analysis other observables.
Alternatively, standard Dirac-Pauli representation of the Dirac
matrices can be used. The four-component analytic solution of the
Dirac equation with the Hamiltonian\begin{equation}\label{dh1}
H=c\mbox{\boldmath$\alpha$}\cdot {\bf p}+\beta
c^{2}-\frac{Z}{\rho}
\end{equation}
in two spatial dimensions has been obtained by Guo et al.
\cite{GY2}. The authors have investigated two \textit{decoupled}
eigenstates of $H$
\begin{eqnarray}\label{p1}
\Psi^{(1)}=\left[ \begin{array} {c}
f_{1}(\rho)e^{i(\mu-1/2)\phi}\\0\\0\\ig_{1}(\rho)e^{i(\mu+1/2)\phi}
\end{array}\right],
\end{eqnarray}
\begin{eqnarray}\label{p2}
\Psi^{(2)}=\left[ \begin{array} {c}
0\\f_{2}(\rho)e^{i(\mu+1/2)\phi}\\ig_{2}(\rho)e^{i(\mu-1/2)\phi}\\0
\end{array}\right],
\end{eqnarray}
where $\mu$ is the eigenvalue of the $z-$component of the total
angular momentum $j_{z}$ defined as:
\begin{equation} j_{z}=l_{z}+\frac{1}{2}\sigma'_{z},
\end{equation}
with
\begin{eqnarray}
\sigma'_{z}=\left[ \begin{array} {cc} \sigma_{z}&0\\0&\sigma_{z}
\end{array}\right],
\end{eqnarray}and
\begin{equation}
l_{z}=-i\partial/\partial\phi.
\end{equation}
In non-relativistic limit the states  (\ref{p1}) and (\ref{p2})
describe an electron with spin \textit{up} and \textit{down},
respectively. Without any additional information about other
conserved quantities the radial amplitudes of these two decoupled
states have to be determined by two different sets of radial
equations \cite{GY2}. It seems a little undesirable, in particular
if we appeal to the three-dimensional theory where only one pair
of radial equations appears for all quantum states.

The purpose of this paper is the analysis of integrals of motion
of the two-dimensional relativistic hydrogen atom and
classification of the states based on \textit{good} quantum
numbers. These goals are achieved by introducing, into Dirac
Hamiltonian, hermitian operators associated with conserved
quantities. Another important aspect of the presented approach is
the possibility of determining all radial functions from one
system of radial equations.

\section{Exact solution for field-free atom}
We have find, that two linearly-independent states (\ref{p1}) and
(\ref{p2}) are the eigenstates of an operator
\begin{equation}\label{dh6}
P=\beta\sigma'_{z},
\end{equation}
which commute with the Hamiltonian (\ref{dh1}). It follows from
(\ref{dh6}) that $P$ is an involution ($P^{2}=1$) and has two
eigenvalues $\eta=\pm 1$. In the non-relativistic limit the
different signs of $\eta$ correspond to states with opposite spin
directions. Moreover, it can be directly verified, by computing
relevant commutators, that beyond operators $j_{z}$ and $P$ there
exists an operator $K$
\begin{equation}\label{k}
K=\beta(\sigma'_{z}l_{z}+\frac{1}{2}),\end{equation} which
commutes with $H$ and both operators $P,j_{z}$. The eigenvalue of
$K$ can be referred to the Dirac quantum number $\kappa$. To
determine a physical meaning of quantum number $\kappa$, let us
derive a relation between $\kappa$ and $\mu$. If we consider the
square of $K$,
\begin{equation}
K^{2}=(\sigma'_{z}l_{z})^{2}+\sigma'_{z}l_{z}+\frac{1}{4}=(l_{z}+\frac{1}{2}\sigma'_{z})^{2},
\end{equation}
we obtain
\begin{equation}
K^{2}=j_{z}^{2},
\end{equation}which means that eigenvalues of $K$ satisfy the relation
\begin{equation}\label{km}
\kappa=\pm\mid\mu\mid.
\end{equation}
We note that operators $K,P$ and $j_{z}$ are not independent. They
fulfil the relation
\begin{equation}\label{kpj}
K=Pj_{z},
\end{equation}
which gives a similar relation for eigenvalues
\begin{equation}\label{kme}
\kappa=\mu\eta.
\end{equation}
It appears from (\ref{kme}) that the upper and lower signs in
(\ref{km}) distinguish between two different physical situations,
when, in non-relativistic limit, spin is parallel or antiparallel to the total angular momentum.

Now we are in a position to solve the 2-D Coulomb problem in
strict analogy to 3-D case and perform the classification of
states free of non-relativistic quantum numbers. To this end we
introduce to the Dirac equation quantum numbers associated with
the complete set of commuting operators $(H, K, j_{z})$.

In atomic units and polar coordinates $(\rho,\phi)$ Hamiltonian
(\ref{dh1}) can be written in the form
\begin{equation}\label{dh}
H=c(\alpha_{\rho}p_{\rho}+\frac{i}{\rho}\alpha_{\rho}\sigma'_{z}l_{z})+\beta
c^{2}-\frac{Z}{\rho},
\end{equation}
where\begin{equation}\label{dh3} \alpha_{\rho}=\left[
\begin{array} {cc} 0&\mbox{\boldmath$\sigma$}\cdot \hat{\mbox{\boldmath$\rho$}}
\\\mbox{\boldmath$\sigma$}\cdot \hat{\mbox{\boldmath$\rho$}}&0
\end{array}\right],\hspace{0.2cm}\mbox{\boldmath$\sigma$}\cdot \hat{\mbox{\boldmath$\rho$}}=\left[
\begin{array} {cc} 0&e^{-i\phi}
\\e^{i\phi}&0
\end{array}\right],
\end{equation}
and
\begin{equation}\label{dh5}
p_{\rho}=-i\frac{\partial}{\partial\rho}.\end{equation} Taking
into account the algebraic properties of matrices $\alpha$,
$\beta$ and definition (\ref{k}) of $K$ we obtain
\begin{equation}
H=c(\alpha_{\rho}\pi_{\rho}+\frac{i}{\rho}\alpha_{\rho}\beta
K)+\beta c^{2}-\frac{Z}{\rho},
\end{equation}where the radial momentum $\pi_{\rho}$ is defined
as
\begin{equation}
\pi_{\rho}=-i(\frac{\partial}{\partial\rho}+\frac{1}{2\rho}).
\end{equation}

In the representation in which operators $H, j_{z}$ and $K$ are
diagonal, energy levels are determined by radial part of the wave
function only. The pertinent radial Dirac equation takes the form
\begin{equation}\label{rd}
[c(\alpha_{\rho}\pi_{\rho}+\frac{i\kappa}{\rho}\alpha_{\rho}\beta)+\beta
c^{2}-\frac{Z}{\rho}]R=WR,
\end{equation}where $W$ denotes energy and $Z$ is the charge of the Coulomb field.
Since $\alpha_{\rho}$ and $\beta$ fulfil relations
\begin{equation}
\alpha_{\rho}^{2}=\beta^{2}=1,\hspace{0.2cm}\alpha_{\rho}\beta+\beta\alpha_{\rho}=0
\end{equation}they can be represented by two dimensional hermitian
matrices \begin{eqnarray}\label{dm} \alpha_{\rho}=\left[
\begin{array} {cc}
0&-i\\i&0
\end{array}\right],\hspace{0.2cm}\beta=\left[
\begin{array} {cc}
1&0\\0&-1
\end{array}\right].\end{eqnarray}
According to Eq. (\ref{dm}) the radial function $R$ has two
component, which for convenience we take in the form
\begin{eqnarray}\label{rf}
R(\rho)=\frac{1}{\rho^{1/2}}\left[ \begin{array} {c}
F(\rho)\\G(\rho)Z/c
\end{array}\right].
\end{eqnarray}Introducing two new variables
\begin{equation}\label{nv}
r=Z\rho,\hspace{0.2cm}E=\frac{W-c^{2}}{Z^{2}}
\end{equation}and substituting (\ref{dm}) into (\ref{rd}) leads to the wave equation for the
electron moving in the two-dimensional Coulomb field in the form
of the pair of $\kappa$-dependent radial equations:
\begin{equation}\label{de1}
\frac{dG}{dr}+\frac{\kappa}{r}G+(\frac{1}{r}+E)F=0,
\end{equation}
\begin{equation}\label{de2}
\frac{dF}{dr}-\frac{\kappa}{r}F-[\lambda(\frac{1}{r}+E)+2]G=0,
\end{equation}where $\lambda=(Z/c)^{2}$.

Since asymptotic solutions of Eqs. (\ref{de1}), (\ref{de2}) decay
exponentially we try to find the solution of radial equations
(\ref{de1}) and (\ref{de2}) in the form
\begin{equation}\label{ps}
F=r^{\gamma}e^{-\alpha
r}\sum_{i=0}^{\infty}a_{i}r^{i},\hspace{0.5cm}G=r^{\gamma}e^{-\alpha
r}\sum_{i=0}^{\infty}b_{i}r^{i}.
\end{equation}where
\begin{equation}\label{alp}
\alpha=\sqrt{-E(2+\lambda E)}.
\end{equation}
Substituting expansions (\ref{ps}) into equations (\ref{de1}),
(\ref{de2}) we obtain the linear relations between the expansion
coefficients\begin{equation}\label{le1}
(i+\gamma-\kappa)a_{i}-\lambda b_{i}=\alpha a_{i-1}+(2+\lambda
E)b_{i-1},
\end{equation}
\begin{equation}\label{le2}
a_{i}+(i+\gamma+\kappa)b_{i}=-Ea_{i-1}+\alpha b_{i-1}.
\end{equation}
 For $i=0$, we have
\begin{equation}
(\gamma-\kappa)a_{0}-\lambda
b_{0}=0,\hspace{0.2cm}a_{0}+(\gamma+\kappa)b_{0}=0.
\end{equation}Since both $a_{0}$ and $b_{0}$ are different from zero, the secular determinant must vanish, which leads to
\begin{equation}\label{gam1}
\gamma=\pm\sqrt{\kappa^{2}-\lambda},
\end{equation}and\begin{equation}\label{mk1}
b_{0}=-\frac{a_{0}}{\gamma+\kappa}.
\end{equation}
For $i>0$, the expanding coefficients can be calculated
iteratively from the relations
\begin{equation}\label{u1}
b_{i}=\frac{(i+\gamma-\kappa+\alpha/E)w_{i}}{i(i+2\gamma)},
\end{equation}
\begin{equation}\label{u2}
a_{i}=w_{i}-(i+\gamma+\kappa)b_{i},
\end{equation}where
\begin{equation}
w_{i}=-Ea_{i-1}+\alpha b_{i-1}.
\end{equation}
The condition of square integrability of the wave-function allows
only the upper sign in Eq. (\ref{gam1}) and requires the
termination of power series (\ref{ps}) at a some power $n'$
\begin{equation}
a_{n'+1}=0,\hspace{0.2cm}b_{n'+1}=0,\hspace{0.2cm}a_{n'}\neq
0,\hspace{0.2cm}b_{n'}\neq 0,
\end{equation}which leads to the condition
\begin{equation}\label{r1}
E a_{n'}=\alpha b_{n'}.\end{equation}Dividing (\ref{u2}) by
(\ref{u1}) for $i=n'$ and taking into account (\ref{alp}) and
(\ref{r1}) leads to the equation for $E$
\begin{equation}
\sqrt{-E(2+\lambda E)}(n'+\gamma)=1+\lambda E.
\end{equation}Solving this equation and substituting
\begin{equation}\label{gam}
\gamma=\sqrt{\kappa^{2}-\lambda},
\end{equation}we obtain
\begin{equation}\label{enr}
E=\frac{1}{\lambda}\big[(1+\frac{\lambda}{(n'+\sqrt{\kappa^{2}-\lambda})^{2}})^{-1/2}-1\big].
\end{equation}
In order to compare the expression (\ref{enr}) with
non-relativistic one, we define the \textit{principal} quantum
number as follows
\begin{equation}
n=n'+\mid \kappa \mid+1/2.
\end{equation}
Since $n'\geq 0$ we must have $\mid \kappa\mid\leq (n-1/2)$.
However for $n'=0$, the number $\kappa$ must have positive value
only. The absence of the $\kappa<0$ for $n'=0$ follows from
equations (\ref{mk1}) and (\ref{r1}), which both imply
\begin{equation}\label{mk2}
(\gamma+\kappa)>0.
\end{equation}
According to (\ref{gam}) $\gamma$ is a real number smaller than
$\mid\kappa\mid$ and inequality (\ref{mk2}) can be satisfied only
if $\kappa$ is positive. Therefore $\kappa$ must fulfils relation
\begin{equation}\label{dkc} \mid\kappa-\frac{1}{2}\mid\leq n-1.
\end{equation}Finally, energy levels are given through
\begin{equation}\label{el}
E_{n\kappa}=\frac{1}{\lambda}\big[(1+\frac{\lambda}{(n-\mid
\kappa\mid-1/2 +\sqrt{\kappa^{2}-\lambda})^{2}})^{-1/2}-1\big].
\end{equation}
In the non-relativistic limit $(\lambda\rightarrow 0)$ we obtain
\begin{equation}
E_{n}=-\frac{2}{(2n-1)^{2}}.
\end{equation}

The complete spin-space-dependence of wave-functions is given by
\begin{eqnarray}\label{wf1}
\Psi_{n\kappa\mu}(r,\phi)=\left[ \begin{array} {c}
f_{n\kappa}(r)\Omega_{\kappa\mu}(\phi)\\g_{n\kappa}(r)\Omega_{-\kappa\mu}(\phi)
\end{array}\right],
\end{eqnarray}where
\begin{equation}
n=1,2,3,...,
\end{equation}
\begin{equation}
\kappa=1/2,-1/2,3/2,-3/2,...,(n-1/2),
\end{equation}
and
\begin{equation}\label{qnm}
\mu=\pm \kappa.
\end{equation}
The \textit{cylindrical} spinor $\Omega_{\kappa\mu}(\phi)$ is
defined as
\begin{eqnarray}\label{wf1}
\Omega_{\kappa\mu}(\phi)=\left[ \begin{array} {c}
\frac{\kappa+\mu}{2\mu}e^{i(\mu-1/2)\phi}\\\frac{-\kappa+\mu}{2\mu}e^{i(\mu+1/2)\phi}
\end{array}\right],
\end{eqnarray}
and radial amplitudes have the form
\begin{equation}
f_{n\kappa}(r)=r^{\gamma-1/2}e^{-\alpha r}\sum_{i=0}^{n-\mid
\kappa \mid-1/2}a_{i}r^{i},
\end{equation}
\begin{equation}
g_{n\kappa}(r)=r^{\gamma-1/2}e^{-\alpha r}\sum_{i=0}^{n-\mid
\kappa\mid-1/2}b_{i}r^{i},
\end{equation}
with coefficients determined by relations (\ref{mk1})-(\ref{u2}).

Alternatively, in a similar way as in 3-D case, the solution of
radial equations (\ref{de1}) and (\ref{de2}) may be expressed in
terms of confluent hypergeometric functions
\begin{equation}\label{fh1}
F(r)=r^{\gamma}e^{-\alpha
r}[(\kappa+\frac{1}{\alpha})F_{1}(r)-n'F_{2}(r)],
\end{equation}
\begin{equation}\label{fh2}
G(r)=\frac{E}{\alpha}r^{\gamma}e^{-\alpha
r}[(\kappa+\frac{1}{\alpha})F_{1}(r)+n'F_{2}(r)],
\end{equation}where
\begin{equation}\label{sh1}
F_{1}(r)={}_{1}F_{1}(-n',2\gamma+1;2\alpha r),
\end{equation}
\begin{equation}\label{sh2}
F_{2}(r)={}_{1}F_{1}(1-n',2\gamma+1;2\alpha r).
\end{equation}

In order to introduce the classification scheme based on
spectroscopic notation it is useful to define an \textit{orbital}
quantum number as
\begin{equation}
l=\mid \kappa-\frac{1}{2}\mid.
\end{equation}Note that this quantity according to (\ref{dkc}) satisfies the
inequality $l\leq (n-1)$.

In Table \ref{t1} we display, as an example, the lowest few states
with principal quantum numbers $n=1,2,3$. We can see that for a
given n, the states with the same $\mid \kappa\mid$ are degenerate and
the lowest energy corresponds to minimal value of $\mid \kappa\mid$.
According to (\ref{qnm}), for each value of $\kappa$ there are
two possible values of $\mu$. Therefore, the
degree of degeneracy of the $n\kappa$-th energy level is 2 for
$\kappa=(n-1/2)$ and 4 for $\mid\kappa\mid<(n-1/2)$, respectively.
It is worth to point out that, in a contrast to the
three-dimensional case, due to the equality $\mid \kappa\mid=\mid
\mu\mid$, the states with $\mid \mu\mid<\mid \kappa\mid$ do not
occur.

\section{Linear Paschen-Back effect}

Let us consider now the Dirac Hamiltonian describing transversal
motion of an electron around a fixed center of Coulomb field with
charge $Z$ and in a static uniform magnetic field. In atomic units
the relativistic Hamiltonian can be written in the form
\begin{equation}
H=c\mbox{\boldmath$\alpha$}\cdot ({\bf p}+{\bf A})+\beta
c^{2}-\frac{Z}{\rho}.
\end{equation}
Taking into account the standard four-dimensional Dirac-Pauli
representation of the Dirac matrices and the vector potential
${\bf A}={\bf B}\times\mbox{\boldmath$\rho$}/2$, for ${\bf
B}=B\hat{z}$ perpendicular to the plane of transversal motion of
the electron, we can write
\begin{equation}
\mbox{\boldmath$\alpha$}\cdot ({\bf p}+{\bf
A})=\alpha_{\rho}p_{\rho}+i\alpha_{\rho}\sigma'_{z}(\frac{l_{z}}{\rho}+\frac{B\rho}{2}).
\end{equation} Introducing into Hamiltonian $H$ both operators $P$
and $K$, defined in a previous section, we obtain
\begin{equation}
H=c(\alpha_{\rho}\pi_{\rho}+\frac{i}{\rho}\alpha_{\rho}\beta
K+\frac{i}{2}B\rho\alpha_{\rho}\beta P)+\beta c^{2}-\frac{Z}{\rho}
\end{equation} and appropriate radial Dirac equation in the form
\begin{equation}\label{rdb}
[c(\alpha_{\rho}\pi_{\rho}+\frac{i\kappa}{\rho}\alpha_{\rho}\beta+\frac{i\eta
B}{2}\rho\alpha_{\rho}\beta)+\beta c^{2}-\frac{Z}{\rho}]R=WR.
\end{equation}

Substituting (\ref{dm}) into (\ref{rdb}) and changing variables
according to (\ref{rf}) and (\ref{nv}) and introducing a new
variable $B:=B/Z^{2}$ leads to the wave equation, for the electron
moving in the superposition of the two-dimensional Coulomb field
and constant homogeneous magnetic field, in the form of the pair
of $\kappa$- and $\eta$-dependent radial equations
\begin{equation}\label{br1}
\frac{dG}{dr}+(\frac{\kappa}{r}+\frac{1}{2}\eta B r
)G+(\frac{1}{r}+E)F=0,
\end{equation}
\begin{equation}\label{br2}
\frac{dF}{dr}-(\frac{\kappa}{r}+\frac{1}{2}\eta B
r)F-[\lambda(\frac{1}{r}+E)+2]G=0.
\end{equation}
The complete spin-space description of eigenfunctions is the same
as for the field-free atom. The only difference is in dependence
of radial functions on both quantum numbers $\kappa$ and $\mu$. In
consequence, magnetic energy shift may depend on the symmetry of
the states. We investigate this problem in the first-order
perturbation approach. In order to applying the perturbation
formalism we rewrite Eqs. (\ref{br1}) and (\ref{br2}) in $2\times
2$ matrix form
\begin{equation}
(h^{(0)}+Bh^{(1)}-ES)\Phi=0,
\end{equation}
where
\begin{eqnarray}
h^{(0)}=\left[ \begin{array} {cc} -1/2 &
-d/dr-\kappa/r\\d/dr-\kappa/r & -(2+\lambda/r)
\end{array}\right],
\end{eqnarray}
\begin{eqnarray}
h^{(1)}=\left[ \begin{array} {cc} 0 & -\eta r/2\\-\eta r/2 & 0
\end{array}\right],
\end{eqnarray}
\begin{eqnarray}
S=\left[ \begin{array} {cc} 1 & 0\\0 & \lambda
\end{array}\right],
\end{eqnarray}
\begin{eqnarray}
\Phi(r)=\left[ \begin{array} {c} F(r)\\G(r)
\end{array}\right].
\end{eqnarray}
Perturbation expansions for energy and wave-function
\begin{equation}
E=\sum_{i=0}^{\infty}E^{(i)}B^{i},\hspace{0.2cm}\Phi=\sum_{i=0}^{\infty}\Phi^{(i)}B^{i}
\end{equation}
lead to the following perturbation equations
\begin{equation}\label{peq}
h^{(0)}\Phi^{(n)}+h^{(1)}\Phi^{(n-1)}-\sum_{i=0}^{n}E^{(i)}S\Phi^{(n-i)}=0.
\end{equation}
The zero-order equation $(n=0)$
\begin{equation}
(h^{(0)}-E^{(0)}S)\Phi^{(0)}=0
\end{equation}
is equivalent to the system of radial equations (\ref{de1}) and
(\ref{de2}) for field-free atom. Under the condition of
orthogonality
\begin{equation}
\langle\Phi^{(0)},S\Phi^{(i)}\rangle=0,
\end{equation}
for $i>0$, the $n$-th order energy correction can be written in
the form
\begin{equation}
E^{(n)}=\frac{\langle\Phi^{(0)},h^{(1)}\Phi^{(n-1)}\rangle}{\langle\Phi^{(0)},S\Phi^{(0)}\rangle}.
\end{equation}
The calculation of the first-order energy correction
\begin{equation}
E^{(1)}=-\eta\frac{\langle F,rG\rangle}{\langle
F,F\rangle+\lambda\langle G,G\rangle}
\end{equation}can be performed in closed form using radial
functions (\ref{fh1}) and (\ref{fh2}). For $n'=0$
$(\kappa=\mid\mu\mid)$, we obtain
\begin{equation}\label{ec12}
E^{(1)}=\frac{\mu}{4\kappa}(2\gamma+1).
\end{equation}
Taking into account that hypergeometric functions (\ref{sh1}) and
(\ref{sh2}) depend only on $\mid \kappa\mid$, the dependence of
functions $F$ and $G$ on $\kappa$ is due to the factor
$(\kappa+1/2)$ in front of $F_{1}$ in (\ref{fh1}) and (\ref{fh2}).
It means that for $n'>0$ the first-order energy corrections have
the general form
\begin{equation}\label{ec4}
E^{(1)}=-\eta\frac{a\kappa+b}{c\kappa+d}=\mu
A_{1}+\frac{\kappa}{\mu}A_{2},
\end{equation}where
\begin{equation}\label{ec14}
A_{1}=\frac{ad-bc}{\kappa^{2}c^{2}-d^{2}},\hspace{0.2cm}A_{2}=\frac{bd-\kappa^{2}ac}{\kappa^{2}c^{2}-d^{2}}.
\end{equation}
Radial integrals a,b,c,d depend only on $\kappa^{2}$ and $n'$ and
are given in Appendix A. It appears from Eqs. (\ref{ec12}) and
(\ref{ec14}) that the $\kappa$- and $\mu$-degeneracy of field-free
levels is completely removed by external magnetic field. Table
\ref{t2} lists the first-order energy shifts of levels
corresponding to states with $n=1,2,3$, presented in Table
\ref{t1}. We can see that linear corrections $E^{(1)}$ essentially
depend on the sign of both quantum numbers $\kappa$ and $\mu$.
Calculating the non-relativistic limit of (\ref{ec12}) and
(\ref{ec4}) we obtain the values of energy corrections
$E^{(1)}_{N}=\lim_{\lambda\rightarrow 0}E^{(1)}$, which consist
with the non-relativistic result
\begin{equation}\label{nrec}
E_{N}^{(1)}=\frac{1}{2}(m+2m_{s}),
\end{equation}
where $m$ and $m_{s}$ are the eigenvalues of $l_{z}$ and $s_{z}=\frac{1}{2}\sigma_{z}$, respectively.
The non-relativistic formula (\ref{nrec}) may be derived from the Schr\"odinger-Pauli
equation in a similar way as in the 3-D case (see for example \cite{LND}).\\

\section{Concluding Remarks}
The exact solution of 2-D hydrogen problem which have been
presented in this paper is consistent with fundamental principles
of  quantum mechanics. Due to introducing \textit{good} quantum
numbers into wave equation, the problem of energy spectrum is
solved exactly in close analogy to the 3-D case. Moreover, the
wave-functions are classified according to the complete set of
constants of motion and determined by one $\kappa$-dependent
system of radial equations. Although, the energy spectrum of
field-free central problem depends only on
$\mid\kappa\mid=\mid\mu\mid$, the full $\kappa$- and
$\mu$-dependence appears when the external magnetic field is
applied.

\appendix\section{}
Radial integrals occurring in the expression of the first-order
magnetic energy correction (\ref{ec4}) are defined as follows
\begin{equation}
a=\frac{2E}{\alpha^{2}}K_{1},\hspace{0.2cm}b=\frac{E}{\alpha}[(\kappa^{2}+\frac{1}{\alpha^{2}})K_{1}-{n'}^{2}K_{2}],
\end{equation}
\begin{equation}
c=\frac{2}{\alpha}(1+\frac{\lambda
E^{2}}{\alpha^{2}})I_{1}+2n'(\frac{\lambda
E^{2}}{\alpha^{2}}-1)I_{12},
\end{equation}
\beqn d=(1+\frac{\lambda
E^{2}}{\alpha^{2}})[(\kappa^{2}+\frac{1}{\alpha^{2}})I_{1}+{n'}^{2}I_{2}]\nonumber\\+\frac{2n'}{\alpha}(\frac{\lambda
E^{2}}{\alpha^{2}}-1)I_{12}, \eeqn where
\begin{equation}
K_{i}=\int_{0}^{\infty}r^{2\gamma+1}e^{-2\alpha r}F_{i}^{2}dr,
\end{equation}
\begin{equation}
I_{i}=\int_{0}^{\infty}r^{2\gamma}e^{-2\alpha r}F_{i}^{2}dr,
\end{equation}where $i=1,2$ and
\begin{equation}
I_{12}=\int_{0}^{\infty}r^{2\gamma}e^{-2\alpha r}F_{1}F_{2}dr.
\end{equation}

\begin{table*}
\caption{Relativistic quantum numbers, spectroscopic notation and
energies for bound states with $n=1,2,3$  for the two-dimensional
hydrogen atom. The values of energy are computed with
$c=137.03599976$ \cite{CD}. }\label{t1}
\begin{ruledtabular}
\begin{tabular}{cccccc}
$n$ & $n'=n-\mid \kappa\mid-1/2$ & $\kappa$ & $l=\mid \kappa-\frac{1}{2}\mid$ & Notation & Energy\\
\hline
$1$ & $0$ &   $1/2$ & $0$ & $1s_{1/2}$ & $-2.000106514052$\\
$$ &  $$ & $$ & $$ & $$ & $$\\
$2$ & $1$ & $1/2$ & $0$ & $2s_{1/2}$ & $-0.222234057055$\\
$2$ & $1$  & $-1/2$ & $1$ & $2p_{1/2}$ & $-0.222234057055$\\
$2$ & $0$  & $3/2$ & $1$ & $2p_{3/2}$ & $-0.222223537086$\\
$$ & $$  & $$ & $$ & $$ & $$\\
$3$ & $2$  & $1/2$ & $0$ & $3s_{1/2}$ & $-0.080002897124$\\
$3$ &  $2$  & $-1/2$ & $1$ & $3p_{1/2}$ & $-0.080002897124$\\
$3$ & $1$  & $3/2$ & $1$ & $3p_{3/2}$ & $-0.080000624824$\\
$3$ & $1$  & $-3/2$ & $2$ & $3d_{3/2}$ & $-0.080000624824$\\
$3$ & $0$ & $5/2$ & $2$ & $3d_{5/2}$ & $-0.080000170405$\\
\end{tabular}
\end{ruledtabular}
\end{table*}

\begin{table}
\caption{First-order magnetic corrections $E^{(1)}$ to the
energies for states with principal quantum numbers $n=1,2,3$ of
the two-dimensional relativistic hydrogen atom. The upper and
lower signs in the front of $E^{(1)}$ are referred to the two
values of $\mu=\pm\mid \kappa\mid$, respectively. The numbers in
brackets are the powers of $10$ by which the entries are
multiplied. In the column 5 nonrelativistic values of linear
magnetic corrections are given.}\label{t2}
\begin{ruledtabular}
\begin{tabular}{ccccc}
$n'$ & $\kappa$ & State & $E^{(1)}$ & $E_{N}^{(1)}$\\
\hline
$0$ & $1/2$ & $1s_{1/2}$ & $\pm 0.49997337$ & $\pm 0.5$\\
$ $ & $3/2$ & $2p_{3/2}$ & $\pm 0.99999112$ & $\pm 1$\\
$ $ & $5/2$ & $3d_{5/2}$ & $\pm 1.49999467$ & $\pm 1.5$\\
$$ & $$ & $$ & $$\\
$1$ & $1/2$ & $2s_{1/2}$ & $\pm 0.49999704$ & $\pm 0.5$\\
$ $ & $-1/2$ & $2p_{1/2}$ & $\mp 2.9586[-6]$ & $0$\\
$ $ & $3/2$ & $3p_{3/2}$ & $\pm 0.99999680$ & $\pm 1$\\
$ $ & $-3/2$ & $3d_{3/2}$ & $\pm 0.49999680$ & $\pm 0.5$\\
$$ & $$ & $$ & $$\\
$2$ &  $1/2$ & $3s_{1/2}$ & $\pm 0.49999899$ & $\pm 0.5$\\
$ $ & $-1/2$ & $3p_{1/2}$ & $\mp 1.0651[-6]$ & $0$\\
\end{tabular}
\end{ruledtabular}
\end{table}

\end{document}